\documentclass[11pt]{article}

\usepackage[final]{acl}
\usepackage{tikz}
\usetikzlibrary{arrows.meta,positioning,calc}
\usepackage{times}
\usepackage{latexsym}
\usepackage{listings}
\usepackage[T1]{fontenc}
\usepackage[utf8]{inputenc}
\usepackage{amsmath}
\usepackage{microtype}

\usepackage{inconsolata}

\usepackage{graphicx}

\title{Sifei at SemEval-2026 Task 8: Hybrid Retrieval and Query Rewriting for Multi-Turn RAG}

\author{Sifei Meng \\
  Faculty of Computer Science \\
  HSE University \\
  \texttt{smen@edu.hse.ru} \\\And
  Dmitry Ilvovsky \\
  Faculty of Computer Science \\
  HSE University \\
  \texttt{dilvovsky@hse.ru} \\}

\begin{document}
\maketitle
\begin{abstract}
Multi-turn retrieval-augmented generation (RAG) is challenging due to evolving user intent, conversational noise, and strict context limits. We propose a training-free, hybrid retrieval pipeline for SemEval-2026 Task~8 that combines dense and sparse retrieval with controlled query rewriting and cross-encoder reranking. On the official test set of Task~A, our system achieves 0.5453 nDCG@5, ranking third among 38 teams and outperforming the strongest baseline (0.4795). For Task~C, we reuse the documents retrieved for Task~A and apply a lightweight generation pipeline guided by the official prompt, achieving 0.5312 (the harmonic mean of relevance and faithfulness) and ranking 15th out of 29 teams. All retrieval components are open-source, while rewriting and generation rely on LLM APIs. Code and scripts are available on GitHub\footnote{\url{https://github.com/mengsifei/Multi\_turn\_RAG}}.
\end{abstract}

\section{Introduction}
Multi-turn retrieval-augmented generation (RAG) is central to real-world chat assistants. While many benchmarks focus on single-turn queries and underrepresent multi-turn phenomena, the English MTRAG benchmark \cite{katsis2024mtrag} captures contextual shifts with human-generated responses, showing that even strong RAG systems degrade on later turns and context-dependent queries.

SemEval-2026 Task~8 (MTRAGEval) ~\cite{Rosenthal2026MTRAGEval} evaluates multi-turn RAG in three subtasks. Task~A focuses on retrieval, where systems rank the top-10 relevant passages. Task~B evaluates response quality given gold relevant passages. Task~C is a RAG setting: systems retrieve up to 10 passages from the corpus and generate a grounded response. We participate in Task~A and Task~C because together they form an end-to-end RAG system.

Our Task~A system is a training-free hybrid pipeline that combines dense and sparse retrievers with corpus cleaning, conversational query rewriting, and reranking. We sweep cleaning strategies, dense backbones, context-window and chunking settings, sparse variants, rerankers, and fusion methods, finding that (i) strict cleaning improves dense retrieval, with Jasper-Token-Compression-600M performing best, (ii) SPLADE-v3 is the strongest sparse component, and (iii) reciprocal rank fusion with a BGE reranker (v2-m3) and an $\alpha$-rescoring yields the best nDCG@5. The final system reaches 0.5453 nDCG@5 on the official test set~\cite{rosenthal2026mtragunbenchmarkopenchallenges}, ranking 3rd out of 38 teams in Task~A. For Task~C, we reuse the retriever and conversational query reconstruction within the official generation framework, achieving 0.5312 and ranking 15th out of 29 teams.

Our contributions are threefold.
(1) We propose a training-free hybrid multi-turn retrieval pipeline that uses open-source retrieval and reranking components and LLM APIs for query rewriting, ranking third in Task~A.
(2) We provide systematic ablations and analysis for Task~A to identify the design choices that yield the largest gains.
(3) We find that concatenating the last turn with a controlled standalone rewrite consistently improves retrieval and generation at no additional training cost.

\section{Background}

\subsection{MTRAGEval Task Setup}
MTRAGEval is built on the MTRAG benchmark, consisting of human-generated multi-turn conversations grounded in four domain corpora: CLAPNQ, FiQA, Govt, and Cloud. The released benchmark contains 110 conversations with 7.7 turns per conversation on average, totaling 842 turn-level tasks; after filtering unanswerable queries, we obtain 777 tasks in our experiments. Given a dialogue history ending with the final user query, \textbf{Task~A} requires an ordered list of $k{=}10$ passages from the corresponding corpus, evaluated by nDCG@5 on the answerable subset. \textbf{Task~C} uses retrieved passages to generate the final assistant response and ranks systems by the harmonic mean of multiple metrics measuring response quality and faithfulness to evidence \cite{katsis2024mtrag}. Following the official release, the organizers provide each corpus as pre-segmented passages (512 tokens with 100-token overlap), and evaluation is performed at the same passage level using these passage IDs. We therefore reuse the provided passage segmentation and passage IDs.

\subsection{Related Work}

Prior work on retrieval-augmented generation (RAG) highlights several recurring components for improving retrieval robustness and ranking quality.

\textbf{Preprocessing and chunking.} Several studies \cite{merola2025reconstructingcontextevaluatingadvanced,zhao2025metachunkinglearningtextsegmentation,bansal2024hybridrag} show that corpus refinement and contextual chunking can improve retrieval alignment.

\textbf{Query reformulation in multi-turn conversations.} For ambiguous or non-standalone queries, query rewriting and keyword-based reformulation are effective \cite{katsis2024mtrag,Chen2025,pattnayak2025hybridairesponsivemultiturn}; concatenating a hypothetical document with the user query is also beneficial \cite{wang2024searchingbestpracticesretrievalaugmented}.

\textbf{Hybrid retrieval to increase recall.} Hybrid pipelines combine sparse and dense retrievers with fusion methods such as reciprocal rank fusion, which consistently improves retrieval quality \cite{kalra2025hyparaghybridparameteradaptive}; hybrid retrieval performance is also sensitive to fusion hyperparameters \cite{wang2024searchingbestpracticesretrievalaugmented}.

\textbf{Reranking to increase precision.} Reranking consistently improves nDCG in competitive RAG settings, and ablations identify it as a major contributor to final ranking quality \cite{lee2024multireranker}.

We evaluate these techniques on MTRAGEval and provide ablations to quantify their impact.

\section{System Overview}
% Preamble:
% \usepackage{tikz}
% \usetikzlibrary{arrows.meta,positioning,calc}

\begin{figure}[t]
\centering
\resizebox{0.75\linewidth}{!}{%
\begin{minipage}{0.48\textwidth}
\centering
\begin{tikzpicture}[
  font=\small,
  node distance=6mm and 14mm,
  block/.style={
    draw=black!60, rounded corners=2pt, fill=black!3,
    inner xsep=6pt, inner ysep=5pt, align=center
  },
  arrow/.style={-Latex, thick},
  dashedarrow/.style={-Latex, thick, dashed},
  scale=0.92, transform shape
]

% Top: query on left, corpus on right
\node[block] (qset)   {\textit{Query dataset}};
\node[block, right=28mm of qset] (cset) {\textit{Corpus dataset}};

\node[block, below=of qset] (qrec) {\textit{Query reconstruction}};
\node[block, below=of cset] (cclean) {\textit{Corpus cleaning}};

% Retrieval row
\node[block, below=12mm of qrec] (dense)  {\textit{Dense retrieval}\\(\textit{top-200})};
\node[block, below=12mm of cclean] (sparse) {\textit{Sparse retrieval}\\(\textit{top-200})};

\node[block] (rrf) at ($(dense)!0.5!(sparse)$) {\textit{RRF fusion}\\(\textit{top-200})};

% Downstream
\node[block, below=10mm of rrf] (rerank)
  {\textit{Cross-encoder reranking}\\(\textit{rerank top-100, keep top-100})};
\node[block, below=of rerank] (alpha) {\textit{$\alpha$-rescoring}\\(\textit{keep top-100})};
\node[block, below=of alpha] (out) {\textit{Top-$k$ output}\\(\textit{top-10})};

% Vertical arrows
\draw[arrow] (qset) -- (qrec);
\draw[arrow] (cset) -- (cclean);

% Both sources feed both retrievers (no curved arrows)
\draw[arrow] (qrec) -- (dense);                 % query -> dense
\draw[arrow] (cclean) -- (sparse);              % corpus -> sparse

\draw[arrow] (qrec.south east) -- (sparse.north west);     % query -> sparse
\draw[arrow] (cclean.south west) -- (dense.north east);    % corpus -> dense

% Retrieval -> fusion from left/right
\draw[arrow] (dense.east) -- (rrf.west);
\draw[arrow] (sparse.west) -- (rrf.east);

% Downstream
\draw[arrow] (rrf) -- (rerank);
\draw[arrow] (rerank) -- (alpha);
\draw[arrow] (alpha) -- (out);

% Fallback (orthogonal dashed, no curves)
\path (sparse.east) ++(10mm,0) coordinate (fbx); % routing x
\draw[dashedarrow] (sparse.east) -- (fbx) |- (cset.east)
  node[pos=0.35, right, align=left, text width=1.0cm]
  {\scriptsize Fallback check};

\end{tikzpicture}
\caption{Hybrid retrieval and reranking workflow.}
\label{fig:workflow}
\end{minipage}
}
\end{figure}
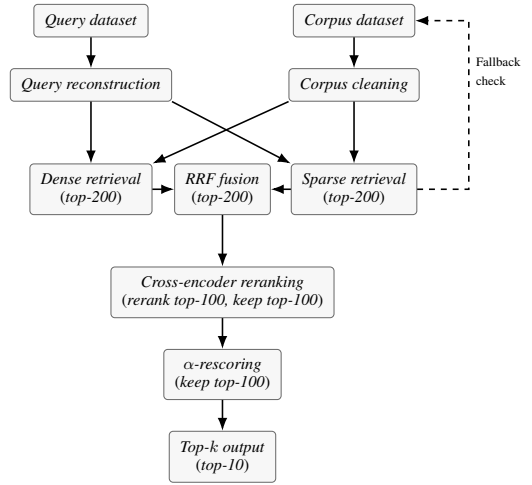

Our submissions follow a multi-stage hybrid retrieval pipeline for Task~A, and reuse the same retriever for Task~C generation. The overall retriever workflow is shown in Figure \ref{fig:workflow}.

\subsection{Query reconstruction}
\label{sec:qr}
Given a multi-turn dialogue, we construct the retrieval query by concatenating the last user turn with a custom standalone rewrite generated by GPT-5.2 (prompt provided in Appendix~\ref{app:rewrite-prompt} and fixed before official evaluation). The last turn preserves surface phrasing, while the rewrite resolves implicit constraints and reduces ambiguity. The final query is:
\[
q = \texttt{last\_turn} \; \Vert \; \texttt{custom rewritten query}.
\]
This concatenation consistently outperforms either component alone and is used for both retrieval (Task~A) and generation (Task~C). All rewritten queries are generated once, cached, and not modified after official test submission; the files are included in our submission artifacts.

\subsubsection{Analysis of Conversational Rewriting}
On the MTRAG training split, our rewritten queries achieve lower language-model perplexity than the official rewrites across all four domains (CLAPNQ: 91.60$\rightarrow$77.69; Cloud: 124.88$\rightarrow$105.08; FiQA: 78.42$\rightarrow$60.64; Govt: 91.35$\rightarrow$90.05), suggesting more regular and self-contained standalone formulations. We also observe a small reduction in ambiguous deictic markers (e.g., \emph{this}, \emph{that}, \emph{these}), indicating improved contextual grounding.

\subsection{Corpus preparation and indexing}
To reduce noise in dense retrieval, we build a cleaned corpus by removing non-English passages, near-duplicates (detected via normalized-text hashing, e.g., whitespace and markup canonicalization), ultra-short fragments, and pathological overlong texts, while protecting any qrels-referenced passages from filtering. In hybrid retrieval, we query the cleaned corpus for both dense and sparse components; when cleaning invalidates sparse hits for a given query (i.e., some retrieved passage IDs are absent from the cleaned index), we rerun sparse retrieval on the original corpus for that query to preserve lexical coverage.

\subsection{Stage-1 retrieval: dense and sparse}

We retrieve an initial candidate pool with hybrid retrieval. For dense retrieval, we use Jasper-Token-Compression-600M with 0.5 compression ratio \cite{zhang2025jaspertokencompression600mtechnicalreport}, selected for its strong English MMTEB performance \cite{enevoldsen2025mmtebmassivemultilingualtext}; after sweeping settings, we use a 768-token query/passage budget to retrieve top-200 candidates. For sparse retrieval, we use SPLADE-v3 \cite{lassance2024spladev3newbaselinessplade} to retrieve another top-200, which helps on entity-centric and exact-match queries via stronger lexical anchoring.

\subsection{Hybrid fusion}
We merge dense and sparse ranked lists using Reciprocal Rank Fusion (RRF) with $k{=}10$, producing a single hybrid candidate list of the top-200 fused candidates for downstream reranking. RRF provides a robust combination without requiring score-scale calibration between heterogeneous retrievers. 

\subsection{Cross-encoder reranking and $\alpha$-rescoring}
From the fused list, we rerank the top-100 candidates using the cross-encoder BAAI/bge-reranker-v2-m3~\cite{chen2024bge}, truncating inputs to 1024 tokens. To further calibrate the final ordering, we perform $\alpha$-weighted score fusion between the reranker score and the initial retrieval score:
\[
s(d) = \alpha \cdot s_{\text{retr}}(d) + (1-\alpha)\cdot s_{\text{ce}}(d),
\]
where $\alpha$ is tuned on the released training split only (final $\alpha{=}0.15$) and fixed before official test submission, $s_{\text{retr}}(d)$ is the hybrid stage-1 retrieval score after RRF fusion, and $s_{\text{ce}}(d)$ is the cross-encoder score for candidate $d$. We then output the top-10 passages as the Task~A result.

\subsection{Task C pipeline}
For Task~C, we feed the model the top-5 retrieved passages and the same concatenated query reconstruction from Task~A, which also helps generation. Due to limited development time, we keep the generation module lightweight and close to the official setup.

\section{Experimental Setup}

\subsection{Data usage and evaluation protocol}
We do not fine-tune any model in this work. The dataset is relatively small, which makes stable supervised tuning difficult. We therefore use the entire training dataset for development-time evaluation and follow the official passage-level segmentation and evaluation setup. All hyperparameters (including $\alpha$, fusion parameters, and candidate pool sizes) are tuned on the training dataset only and fixed before official test submission. For Task~A we report nDCG@5 on the answerable subset; for Task~C we report the official harmonic-mean score over response quality and faithfulness metrics \cite{katsis2024mtrag}. Unless otherwise stated, all development-split retrieval results are averaged over three repeated runs; we observed only minor variation across runs.

\subsection{Model selection and hyperparameter sweeps}

We conduct systematic sweeps over corpus cleaning strategies, dense and sparse retrievers, context length, fusion strategies, rerankers, and query reconstruction. 

For dense retrieval, we evaluate multiple open-source embedding models, including BGE-1.5-base\footnote{\scriptsize\url{https://huggingface.co/BAAI/bge-base-en-v1.5}}, Qwen3-Embedding-0.6B\footnote{\scriptsize\url{https://huggingface.co/Qwen/Qwen3-Embedding-0.6B}}, Jasper-Token-Compression-600M\footnote{\scriptsize\url{https://huggingface.co/infgrad/Jasper-Token-Compression-600M}}, gte-Qwen2-1.5B-instruct\footnote{\scriptsize\url{https://huggingface.co/Alibaba-NLP/gte-Qwen2-1.5B-instruct}}, and Yuan-embedding-2.0-en\footnote{\scriptsize\url{https://huggingface.co/IEITYuan/Yuan-embedding-2.0-en}}. Sparse retrieval candidates include BM25 and SPLADE variants (v2-distil\footnote{\scriptsize\url{https://huggingface.co/naver/splade_v2_distil}} and v3\footnote{\scriptsize\url{https://huggingface.co/naver/splade-v3}}). 
For reranking, we consider cross-encoders such as BGE-reranker-v2-m3\footnote{\scriptsize\url{https://huggingface.co/BAAI/bge-reranker-v2-m3}}, all-MiniLM-L6-v2\footnote{\scriptsize\url{https://huggingface.co/sentence-transformers/all-MiniLM-L6-v2}}, and Qwen3-Reranker-0.6B\footnote{\scriptsize\url{https://huggingface.co/Qwen/Qwen3-Reranker-0.6B}}.

We tune retrieval and fusion hyperparameters on the development evaluation, including dense context window size, RRF parameters, and an $\alpha$ sweep for score calibration. 

Our final configuration uses Jasper-Token-Compression-600M for dense retrieval, SPLADE-v3 for sparse retrieval, RRF fusion with tuned $k=10$, BGE-reranker-v2-m3 for cross-encoder reranking, and $\alpha$-rescoring with $\alpha=0.15$.

\subsection{Generation setup}
For Task~C, we reuse the Task~A retriever and provide the top-5 retrieved passages together with the same concatenated query (last turn + standalone rewrite); the full prompt is given in Appendix~\ref{app:taskc-prompt}. Due to GPU constraints, we use the DeepSeek-V3.2 API for answer generation with low-temperature decoding (temperature=0.2, top\_p=1.0, max\_new\_tokens=300). During development, we used DeepSeek-V3.2 as an LLM-as-a-judge via API (temperature=0, max\_new\_tokens=800) for metrics RL\_F and RB\_llm. Official Task~C scores are computed by the organizers using the provided evaluation scripts and the harmonic mean of RL\_F, RB\_llm, and RB\_alg.

\subsection{Implementation and compute}
All experiments in Task~A use a single NVIDIA V100 GPU (32GB). We cache document embeddings for dense retrieval to enable efficient sweeps over fusion and reranking settings.

\section{Results}

\subsection{Official leaderboard performance}
Table~\ref{tab:leaderboard-top3-ours} reports our official results. Our hybrid retriever achieves 0.5453 nDCG@5 on Task~A, ranking 3rd out of 38 teams. On Task~C, our system yields 0.5312, ranking 15th out of 29. The faithfulness score is substantially higher than the relevance scores on the evaluation set (RB\_alg 0.3841, RL\_F 0.7286, RB\_llm 0.5982), suggesting that the retrieved documents provide strong grounding for answer generation.

Compared with the top systems, our method is competitive on retrieval while relying mainly on open-source components. Top Subtask~A systems consistently used conversational query rewriting to produce self-contained queries before retrieval, and two of the top three teams used sparse retrieval. In particular, the winning AILS-NTUA system combined five history-aware rewrites with Elser-v1 retrieval, cohere-rerank-v4.0-pro reranking, and weighted RRF fusion. One factor that may have contributed to this result is the use of Elser-v1, given that the organizers note a possible benchmark bias toward Elser because it was used during data creation \cite{katsis2024mtrag}. In addition, using multiple rewrites may help expose complementary aspects of the information need before fusion. 

For Subtask~C, the top systems, GenAIus and Howard University-AI4PC, employed richer hybrid pipelines with multi-query rewriting, HyDE, lexical--dense retrieval, and reranking. Both also used BM25, which was less effective than SPLADE-v3 in our own sparse-retrieval experiments, suggesting that BM25 may still be useful as a complementary lexical branch inside a stronger hybrid pipeline. Generation was performed in a zero-shot setting with large models such as GPT-4o and Llama-3.3-70B-Instruct.

\begin{table}[t]
\centering
\small
\begin{tabular}{cllc}
\hline
\multicolumn{4}{c}{Subtask A (nDCG@5)} \\
\hline
Rank & Team & Score & Note \\
\hline
1 & AILS-NTUA      & 0.578 & Sparse \\
2 & uva-irlab-conv & 0.548 & Sparse \\
3 & Ours (Sifei)   & 0.545 & Hybrid \\
\hline
\multicolumn{4}{c}{Subtask C (HM)} \\
\hline
Rank & Team & Score & Note \\
\hline
1  & GenAIus       & 0.586 & GPT, ZS \\
2  & HowardU-AI4PC & 0.580 & Llama, ZS \\
3  & 5ting         & 0.560 & GPT, ZS \\
15 & Ours (Sifei)  & 0.531 & DS, ZS \\
\hline
\end{tabular}
\caption{Top-3 leaderboard entries and our submission on MTRAGEval.}
\label{tab:leaderboard-top3-ours}
\end{table}

\subsection{Task A ablation study}

Table~\ref{tab:ablation} shows a progression from recall-oriented improvements to precision-focused refinements.

As a baseline, we use dense-only retrieval with Jasper-Token-Compression-600M at a compression ratio of 0.5 in the last-turn setting, i.e., searching with the raw final user query without rewriting.

Corpus cleaning yields a modest but consistent gain, as dense retrievers are sensitive to templated boilerplate, duplicated blocks, and malformed passages that introduce spurious neighbors. Increasing the dense context window to 768 tokens further improves performance by preserving more entities and constraints in the embedding input. This context length is not tied to the official 512-token segmentation, since our dense encoder relies on a different tokenizer and embedding scheme; empirically, 768 tokens cover over 99\% of passages after preprocessing.

Query rewriting produces a larger improvement, suggesting that many last-turn queries lack sufficient standalone context. Compared with the official rewrites, our custom rewrites have lower language-model perplexity, indicating cleaner and more regular formulations.

Concatenating the last turn with the rewrite outperforms either component alone, likely because it preserves the user’s phrasing while adding missing contextual information. Finally, combining dense retrieval with SPLADE-v3 via RRF substantially increases nDCG by recovering entity-heavy and exact-match evidence that dense retrieval alone may miss. We find that fusion and reranking are sensitive to design choices: small candidate pools limit reranking gains, whereas large pools (e.g., 500–1000) introduce noise that can degrade final ranking quality.

The gap between the train-split ablation result (0.4896 nDCG@5) and the official test result (0.5453) is expected, since the two numbers were obtained on different splits.

\begin{table}[t]
\centering
\small
\setlength{\tabcolsep}{3pt}
\begin{tabular}{p{0.62\linewidth}cc}
\hline
Setting & nDCG@5 & $\Delta$ \\
\hline
Last-turn baseline & 0.3457 & -- \\
+ Cleaned corpus (qrels-protected) & 0.3497 & +0.0040 \\
+ Context Window (768 tokens) & 0.3507 & +0.0010 \\
+ Official rewrite queries & 0.3938 & +0.0431 \\
+ Custom rewrite queries & 0.4144 & +0.0206 \\
+ Concat(last-turn $\Vert$ rewrite) queries & 0.4150 & +0.0006 \\
+ SPLADE-v3 + RRF (k=10) & 0.4668 & +0.0518 \\
+ BGE reranker v2-m3 + $\alpha$-rescoring ($\alpha{=}0.15$) & 0.4896 & +0.0228 \\
\hline
\end{tabular}
\caption{Key ablations for Task~A on the released train split used for pipeline development. All nDCG@5 numbers are averaged over three runs.}
\label{tab:ablation}
\end{table}

\subsection{Task~C Ablation Study}
\label{app:taskc_ablation}

Due to time constraints, we devoted less effort to Task~C and largely followed the official prompt. Still, we observed a practical gain from simply concatenating the rewritten query with the last-turn query. We therefore compare two query formulations for Task~C under the same pipeline and evaluation setting:
(1) \textbf{Rewrite}: use the custom rewritten standalone query, as described in Section~\ref{sec:qr};
(2) \textbf{Concatenation}: concatenate the \emph{raw last user turn} with the rewritten query.
We report RB\_alg, RL\_F, RB\_llm, and their harmonic mean (HM).

\begin{table}[t]
\centering
\small
\setlength{\tabcolsep}{3pt}
\begin{tabular}{lcccc}
\hline
\textbf{Method} & \textbf{HM} $\uparrow$ & \textbf{RB\_alg} $\uparrow$ & \textbf{RL\_F} $\uparrow$ & \textbf{RB\_llm} $\uparrow$ \\
\hline
Query Rewrite & 0.5813 & 0.4415 & 0.8540 & 0.6770 \\
Concatenation & \textbf{0.5922} & \textbf{0.4481} & \textbf{0.8563} & \textbf{0.6910} \\
\hline
$\Delta$ & +0.0109 & +0.0066 & +0.0023 & +0.0140 \\
\hline
\end{tabular}
\caption{Task~C ablation on query formulation (training split). Concatenation improves HM by 1.87\% relative (0.5813 $\rightarrow$ 0.5922), with the largest gain on RB\_llm.}
\label{tab:taskc_ablation_concat_rewrite}
\end{table}

As shown in Table~\ref{tab:taskc_ablation_concat_rewrite}, \textbf{Concatenation} improves all metrics, with the largest gain on \textbf{RB\_llm} (+0.0140). One possible explanation is that rewriting alone can over-compress the query and bias it toward the rewriter’s interpretation. Appending the \textbf{raw last turn} preserves additional cues from the user’s phrasing and anchors the model to the current question. This simple change strengthens both generation relevance and faithfulness, providing a simple and effective improvement for Task~C.

\subsection{Error analysis on the test set (Task A)}
% \label{subsec:error-analysis}
% We analyze our official Task~A test predictions after the release of gold annotations, reporting only aggregate statistics without revealing test data content. The analysis focuses on the answerable subset (285 queries).

\paragraph{Score distribution.}
The score distribution is skewed: 61.1\% of queries achieve nDCG@5 $>$ 0.5, while 16.8\% obtain nDCG@5 $=0$; 2.8\% fall in $(0, 0.2]$ and 19.3\% fall in $(0.2, 0.5]$. Thus, errors concentrate in a relatively small subset of difficult instances.

\paragraph{Stage-level failures.}
For low-score cases (nDCG@5 $<0.2$), we categorize errors into two types: (i) \emph{recall failures}, where no relevant passage appears in the top-10, and (ii) \emph{rerank failures}, where a relevant passage is retrieved but not promoted to the top ranks. Under this diagnostic, 85.7\% of hard cases are recall failures and 14.3\% are rerank failures, indicating that insufficient recall in the final top-10 is the primary bottleneck.

\paragraph{Domain and turn effects.}
Recall failures are most frequent in the Cloud and FiQA domains of the MTRAG benchmark \cite{katsis2024mtrag}, which contain a higher proportion of noisy text, underscoring the importance of effective data cleaning. Hard cases also concentrate in later conversation turns (especially after the fifth turn), indicating that query rewriting remains challenging as dialogue context grows.

In contrast, reranking failures are substantially less common, implying that improving first-stage recall is likely to yield larger gains than further reranker optimization.

\section{Discussion}

\paragraph{Alternative chunking did not help.}
We evaluated several segmentation strategies beyond the official fixed-window passages, including (i) semantic re-chunking on top of the official passages, (ii) document-level chunking with outputs mapped back to the official passage IDs, and (iii) passage packing. None consistently outperformed the official setup. We attribute the degradation mainly to: (i) over-fragmentation into small chunks that lack sufficient context; (ii) evidence being split across multiple units, weakening relevance signals; (iii) a mismatch between chunk-level retrieval scores and passage-level evaluation; and (iv) redundancy introduced by overlap in packing. Together, these factors reduce discriminative power under passage-level ranking.

\paragraph{Reranking and fusion order matter.}
Hybrid retrieval with RRF improves recall but does not always improve top-rank precision (e.g., nDCG@1/3). Cross-encoder reranking helps refine the top results, yet very large candidate pools (e.g., 1000) introduce many near-duplicates and hard negatives, making the final ranking sensitive to small scoring differences and degrading nDCG@5. We find that both the fusion–reranking order and the candidate pool size affect performance. In practice, reranking a moderate pool (top-100) followed by $\alpha$-rescoring ($\alpha{=}0.15$) yields the most stable results.

\paragraph{Keyword-style augmentation underperformed.}
Retrieval-oriented rewrites (e.g., keyword lists or hypothetical-document expansion) consistently hurt performance by distorting intent and over-amplifying rare terms. In contrast, controlled conversational rewriting combined with last-turn concatenation provides consistent gains across domains.

\section{Conclusion}
We presented an end-to-end RAG system for SemEval-2026 Task 8 (MTRAGEval), featuring a multi-stage hybrid retrieval pipeline for Task~A and a lightweight generation pipeline for Task~C. Our retrieval system combines qrels-protected corpus cleaning, conversational query rewriting with last-turn concatenation, dense and sparse retrieval, calibrated fusion, and cross-encoder reranking. Our final submission ranked 3rd in Task~A, achieving 0.5453 nDCG@5. Ablations show that query reconstruction and hybrid retrieval deliver the largest gains, and concatenating the last-turn query with its rewrite consistently improves performance in both Tasks~A and~C at no additional training cost.

Future work will focus on more robust late-turn rewriting and evidence alignment, as well as improving response generation for both answerable and underspecified queries.

\paragraph{Ethical considerations.}
Our system is evaluated on the provided public benchmark and uses only the released conversations and corpora.

\bibliography{custom}

\appendix

\section{Appendix}
\label{sec:appendix}

\subsection{Prompt for query rewriting}
\label{app:rewrite-prompt}

\begin{quote}
\textbf{Purpose:} Rewrite a multi-turn conversational query into a single, clean, standalone English question, following the style of official rewrites.

\textbf{Input:} A conversation history with multiple user turns. The last line is the current user question; earlier lines provide context.

\textbf{Task:} Rewrite the last user question only into a standalone question that can be understood without the conversation history. Use earlier turns only to resolve ambiguity (pronouns, ellipsis, vague references).

\textbf{Requirements:}
\begin{enumerate}
    \item Output exactly one complete English question.
    \item Preserve the original intent and interrogative structure (who/what/where/when).
    \item Do not introduce new concepts; do not merge multiple questions; do not copy the conversation history.
    \item Remove conversational fillers (e.g., ``No, I meant...'', ``Thanks'', etc.).
\end{enumerate}

\textbf{Output format:} Return a JSON object per example: \{\texttt{"task\_id": ... , "collection": ... , "text": "|user|: ..."}\}. Copy \texttt{task\_id} and \texttt{collection} from the input; \texttt{task\_id} has the format \texttt{hash<::>turn}.

\textbf{Example:} 

\textbf{Conversation history:} 
\begin{lstlisting}[basicstyle=\ttfamily\small, breaklines=true] 
> who takes photos of planes in the air
> No, I meant photos in the air. 
\end{lstlisting}

\textbf{Rewrite:} 
\begin{lstlisting}[basicstyle=\ttfamily\small, breaklines=true] 
> Who takes photos of airplanes while they are in the air? 
\end{lstlisting}
\end{quote}

\subsection{Prompt for Task C generation}
\label{app:taskc-prompt}

\begin{quote}

Given the following documents and a user question, generate a concise answer
(less than 150 words) that is grounded in the provided documents.
\begin{enumerate}
    \item Use only information that can be supported by the documents.
    \item You may paraphrase and summarize across documents.
    \item Do NOT quote verbatim unless necessary.
    \item If the documents do not contain enough information to answer the question, say: "I do not have specific information."
\end{enumerate}

[Documents]

[Document 1]

\{context\_1\}

...

[Document 5]

\{context\_5\}

[Question]

last\_turn\_query || rewrite\_query

[Answer]

\end{quote}

\end{document}